# Interactions between spatial and temporal scales in the evolution of dispersal rate

EMMANUEL PARADIS

*School of Biological Sciences, University of East Anglia, Norwich, NR4 7TJ and British Trust for Ornithology, The Nunnery, Thetford, Norfolk, IP24 2PU, UK*

**Summary**

The evolution of dispersal rate is studied with a model of several local populations linked by dispersal. Three dispersal strategies are considered where all, half, or none of the offspring disperse. The spatial scale (number of patches) and the temporal scale (probability of local extinction) of the environment are critical in determining the selective advantage of the different dispersal strategies. The results from the simulations suggest that an interaction between group selection and individual selection results in a different outcome in relation to the spatial and temporal scales of the environment. Such an interaction is able to maintain a polymorphism in dispersal strategies. The maintenance of this polymorphism is also scale-dependent. This study suggests a mechanism for the short-term evolution of dispersal, and provides a testable prediction of this hypothesis, namely that loss of dispersal abilities should be more frequent in spatially more continuous environments, or in temporally more stable environments.

**Introduction**

Dispersal (permanent movement of individuals among local populations) is a ubiquitous characteristic of living beings. Clearly, any species that has not the ability to move in space at one moment or another is greatly exposed to extinction risks. The evolution of dispersal rate (the proportion of dispersing offspring) has received considerable attention from theoreticians (e.g., Johnson and Gaines, 1990; Ludwig and Levin, 1991; McPeek and Holt, 1992; Olivieri et al., 1995) that showed an apparent paradox: some theoretical studies, based on either ecological or population genetic models, showed that dispersal is counter-selected in a spatially heterogeneous but temporally stable environment (Balkau and Feldman, 1973; Teague, 1977; Asmussen, 1983; Hastings, 1983; Holt, 1985). This paradox is not surprising if we look at the long-term fitnesses of the disperser and philopatric strategies. The fact that some offspring disperse towards unfavourable habitats decreases the fitness of the disperser strategy relative to the philopatric one. Only the strategy with no dispersal can persist in the long-term (Morris, 1991). Temporal variability in the environment, however, favors non-null dispersal rates (Kuno, 1981). Competitve interactions also favors dispersal when the environment is temporally constant (Hamilton and May, 1977; Comins et al., 1980; Frank, 1986). It has been suggested by some theoreticians that a combination of spatial and temporal variations is required to select for dispersal (Gadgil, 1971; Roff, 1975; Levin et al., 1984). Furthermore, some observations of polymorphism in dispersal strategies in natural populations suggest that selection against dispersal could be actually important (Harrison, 1980; Venable, 1985; Gouyon and Couvet, 1987; Olivieri et al., 1990; Peroni, 1994; Roff, 1994a; O'Riain et al., 1996). Most theoretical studies on the evolution of dispersal focus on the equilibrium conditions (often by means of ESS techniques), and do not consider the transient behaviours of their model, thus ignoring the time scale of the evolution of dispersal rate. Nevertheless, such a time scale has been suggested to be critical in the maintenance of dispersal polymorphism in some insects (Roff, 1994a).

    In this paper I examine the temporal scale of the evolution of dispersal rate in a variable and heterogeneous environment, and propose a mechanism for the maintenance

of dispersal polymorphism. This theoretical study results in testable predictions on the short-term evolution of dispersal rate.

**Model**

I considered a model where space was structured into several discrete habitat patches. Population growth during reproduction was modelled by the discrete-time exponential logistic model: $N_{it+1} = N_{it} \exp(r_i(1 - N_{it}/K_i))$, where $N_{it}$ is size of population $i$ at time $t$, $r_i$ is the growth rate of population $i$, and $K_i$ is the carrying capacity for population $i$. After breeding, a fraction $f$ (determined genetically) of the offspring dispersed and were then distributed equally in all patches. No mortality during dispersal was assumed. Each generation consisted of a population growth episode and a dispersal episode. The existence of two kinds of habitat was assumed: a high-quality habitat (source habitat) where $r_i > 0$, and a low-quality habitat (sink habitat) where $r_i < 0$. Persistence in sink habitats may be allowed by dispersal, since an isolated population in this kind of habitat declines to extinction. Three genotypes were considered which differed in the fraction of offspring dispersing: $f = 0$, 0.5, or 1, respectively. Each genotype can mutate to any others at a rate of $10^{-4}$. The evolutionary dynamics of this model were studied by numerical simulations. I considered different values of population growth rates in source ($r_s$) and sink habitats ($r_k$), of carrying capacity in both habitats ($K_s$ and $K_k$, respectively), and number of patches of both habitats ($x_s$ and $x_k$, respectively). For a given combination of these parameters simulations were replicated and run with various initial conditions to check the repeatability of the results and the stability of the equilibria observed. Most simulations were run for 2,500 generations; for some combinations of parameters, equilibrium was not reached after 2,500 generations and were run for longer times (up to 200,000 generations). The initial model was set with $r_s = 1$, $r_k = -1$, $K_s = K_k = 1000$, $x_s = x_k = 1$. This two-patch source-sink model was then modified in order to evaluate: (i) the effect of fragmentation of source and/or sink habitats (by varying $x_s$ and/or $x_k$ and keeping total carrying capacity constant), (ii) increase of 'quality' in the sink habitat (by increasing $r_k$ but keeping it negative), (iii) and variation in patch sizes (by varying $K_s$ and/or $K_k$). Temporal variability in the environment was then included in the model in the form of local extinction. Extinctions in each patch occurred with a probability $p_e$ varying between 0.001 and 0.5, and were not temporally or spatially correlated.

Further simulations were made to evaluate how critical are some assumptions on the dynamics of the model. It was assumed that only three dispersal strategies can exist to make the model more tractable but also because this is realistic with respect to some empirical studies that showed, for some organisms, dispersal strategies to have a similar categorization (Gouyon and Couvet, 1987; Roff 1994a). Some simulations were replicated with a fuller range of genotypes; specifically two situations were examined: 11 genotypes with $f$ = 0, 0.1, 0.2, … or 1; and 21 genotypes with $f$ = 0, 0.05, 0.1, 0.15, … or 1. The assumption on the mutation rate was also tested by replicating some simulations with different values for this parameter ($10^{-5}$ or $10^{-6}$). The importance of the assumed genetic mechanism (i.e., three clones with mutations) was evaluated by running simulations with no mutation at all and five individuals of each genotype in each population as initial conditions.

There were no selective values associated with the different genotypes, they grew equally within each population during the population growth process. Differential growth of genotypes was only caused by differential reproductive output in the whole set of populations.

In the present model variations in spatial and temporal scales were modelled by variations in the grain of the environment, and in the rate of local extinction, respectively (Fahrig, 1992). The scale of dispersal did not vary since in all situations the modelled organisms could reach all patches during dispersal; however, the grain of the encountered environment varied when the number of patches did so. This also applies to the temporal scale of the organisms which was modelled by the time step of the simulations; on the other hand, the temporal scale of the environment varied when the rate of local extinction did so.

**Results**

In all simulations with a constant environment ($p_e = 0$) the equilibrium was finally reached when the $f = 0$ genotype was fixed in all populations. However, the convergence towards this equilibrium was markedly different in relation to the spatial structure of the environment. Results below are with five $f = 1$ individuals in a source patch as initial conditions. For the initial model described above, the mean value of $f$ among populations (denoted µ), quickly decreased and was equal to zero after around 200 generations (Fig. 1). A similar result was observed when both source and sink habitats were fragmented ($2 \leq x_s = x_k \leq 50$). When only the source habitat was fragmented ($2 \leq x_s \leq 200$, $x_k = 1$), the convergence towards the equilibrium µ = 0 was slower: the stronger the fragmentation, the slower the convergence (Fig. 1a). When only the sink habitat was fragmented ($x_s = 1$, $2 \leq x_k \leq 200$), the convergence towards the equilibrium was quick. When the rate of population growth in the sink habitat was increased ($-0.5 \leq r_k \leq -0.005$), the convergence towards the equilibrium µ = 0 was critically slower (Fig. 1b). When patch sizes varied ($50 \leq K_s \leq 5000$, $50 \leq K_k \leq 5000$), convergence was always quick.

I examined the interactions between spatial and temporal scales of the environment by including local extinctions in the simulations with $x_s = 1, 5, 20$, or 100. As expected, selection of high dispersal rates occurred in these simulations, the intensity of this selection being stronger when the rate of local extinction $p_e$ increased (Fig. 2). Important differences, however, were apparent with respect to $x_s$. When $x_s$ was large (20 or 100) µ quickly increased with increasing $p_e$ and the $f = 1$ genotype was selected for even for moderate values of $p_e$. For $x_s = 5$, µ was equal to 0.5 for some range of extinction probability ($0.05 \leq p_e \leq 0.2$). Examination of the genotypic frequencies for these simulations shows that the mixed dispersal strategy ($f = 0.5$) was selected for (Figs. 3 and 4). For a relatively high extinction rate ($p_e = 0.4$), two genotypes coexisted (Fig. 5). For a very high extinction rate ($p_e = 0.5$), the pure dispersal strategy ($f = 1$) was fixed. When $x_s = 1$, the situation was similar to the one just previously described for $x_s = 5$ as long as $p_e \leq 0.2$ (Fig. 2). If $p_e$ increased further then there was no transition with a stable polymorphism as previously described, that is for $p_e = 0.25$ the $f = 0.5$ genotype was fixed, and for $p_e = 0.3$ the $f = 1$ genotype was fixed.

Simulations done to evaluate some of the assumptions gave different output for some aspects of the present model. When a larger range of dispersal rates was assumed and there was no temporal variability in the environment, the final result was not changed. i.e. the $f = 0$ genotype was fixed; fixation occurred after longer times (two to three times compared to the same habitat structure) as on average more mutation events were necessary for this genotype to appear in the populations. When local extinctions occurred ($p_e > 0$), the results changed remarkably. In the different situations examined, there was one genotype almost fixed (one or two other genotypes remained at low frequencies) but fixation occurred after considerably more time (between 10,000 and 20,000 generations), and fluctuations in genotypic frequencies still occurred. There was a positive association between $p_e$ and the $f$-value of the fixed genotype, the genotypes persisting at low frequencies had an $f$-value close to that of the fixed genotype (for instance, when the $f = 0.8$ genotype was nearly fixed, the $f = 0.75$ and $f = 0.85$ genotypes persisted at frequencies lower than 0.05).

When the genotypic mutation rate was varied, the only difference with the results described above was that they were observed after a time that was inversely proportional to the mutation rate. When absence of mutation was assumed, no change was noticed relative to the equilibria, the only difference was that they were reached far more quickly (less than 10 generations with the initial model) because the occurrence of the different genotypes was no more conditioned on mutation (note that now the model is fully deterministic when there is no local extinction). Nevertheless, the same differences were observed with respect to the time to reach equilibrium.

**Discussion**

The present study suggests that interactions between spatial and temporal scales are critical in the evolution of dispersal. Two results emerge that seem fairly new: (i) the difference in selective advantage between a disperser and a philopatric strategy is critically affected by the fragmentation of the favourable habitat, and (ii) a polymorphism in dispersal strategy can be maintained provided the favourable habitat is fragmented and the probability of local extinction is correctly tuned.

The first result can be explained by a mix of effects previously observed in models of the evolution of dispersal. When the environment is constant in time and all patches are of equal quality there is a selective advantage for dispersal because this strategy can replace the philopatric one in all sites while the latter can compete only on the sites where it is present (Hamilton and May, 1977). On the other hand, when habitat quality varies among sites then there is a selective disadvantage for dispersal since some offspring disperse in patches of less quality decreasing the long-term fitness of this strategy (Hastings, 1983; Morris, 1991). In the present model with environment constant in time, the selective advantage of the philopatric strategy could be considerably decreased by fragmentation of the favourable habitat, and needs considerably more time to be fixed when this fragmentation is high than in the two-patch model.

The fact that dispersal is selected for when there is temporal variation in the environment is not surprising (Kuno, 1981; Levin et al., 1984). More interesting is that different results were observed depending on the number of patches of source-habitat and the probability of local extinction. How to explain such results? In the present model no selective values were assigned to the different genotypes, and their relative growth is only due to the dynamic properties of the model. It has been argued that the evolution of dispersal is dictated by the opposite forces of individual and group selections (Van Valen, 1971; Craig, 1982; Olivieri et al., 1995). Individual selection favours the genotype that produces the largest number of offspring by differential growth within each site; group selection favours the genotype that produces the largest number of offspring by differential extinction and colonization among sites. One can question whether it is appropriate to invoke group selection for the evolution of dispersal since dispersal is an individual trait and so should evolve in response to individual selection. The controversy about group selection comes from that some authors claimed that it is not necessary to invoke group selection since individual selection can account for the evolution of individual traits in general (Wilson, 1983). Several recent studies, however, showed that group selection can be an effective evolutionary force on individual traits acting in the same direction than individual selection or in an opposite direction (Goodnight et al., 1992; Avilés, 1993). Individual selection and group selection have the same effects, changes in the frequencies of genes or genotypes through time, but with different processes that are described above. Both selective forces are potentially present in a spatially structured population model (and in all real populations) and need to be considered equally.

In my model all genotypes experience the same extinction rate, so dispersal accounts for differences in the intensity of each type of selection. Dispersal is counter-selected by individual selection since it lowers growth within a patch compared to philopatry, but is selected for by group selection since it has a non-null colonization rate. When the number of patches increases there are greater opportunities for group selection to occur as well as when the probability of local extinction increases. For moderate number of patches and moderate extinction probability, individual selection and group selection balance each other so that the mixed strategy (half of the offspring disperse) is selected for. If the extinction probability slightly increases (for instance $p_e = 0.3$) then a polymorphism is maintained only if the source-habitat is fragmented ($x_S = 5$). On the other hand, the pure dispersal strategy is fixed and there is no polymorphism if $x_S = 1$. This seems contradictory with the above statement that selection for dispersal is more intense when the number of patches increases. The contradiction is only apparent since with this value of $p_e$ the probability of simultaneous extinctions of all populations is relatively high if $x_S = 1$ ($0.3^2 = 0.09$) compared to the case with $x_S = 5$ ($0.3^6 = 0.0007$) leading to more opportunities for group selection to operate.

Several theoretical studies on the evolution of dispersal have been published. The originality of the present study is that I considered varying numbers of habitat patches. Roff (1994b) studied a model where dispersal evolved in relation to habitat persistence time, however, he assumed a different genetic mechanism for the transmission of dispersal rate. Roff concluded that the gene coding for a positive dispersal rate, even if recessive, is maintained in any population occupying a habitat with finite persistence time. The gene coding for a null dispersal rate can invade the population but is never fixed (Roff, 1994a, 1994b). The present model goes further than Roff's, and shows that there could be a relation between habitat persistence time and number of patches influencing the selective advantages of the different dispersal strategies.

McPeek and Holt (1992) showed that a polymorphism for dispersal rate could be maintained in a two-patch model when habitat quality varied spatially and, either habitat selection was assumed, or the environment was temporally variable. In this latter situation, however, a strategy with habitat selection was able to invade the system and the polymorphism collapsed. McPeek and Holt (1992) modelled environmental temporal variability by random draws from a bivariate normal distribution. In the present study, the mechanism maintaining polymorphism in dispersal strategy is distinct from the one evidenced by McPeek and Holt (1992) and I showed that it is scale-dependent.

Among the assumptions made in the model that could affect the conclusions, the one relative to no mortality cost of dispersal may be important since many models suggest that this parameter is critical (Johnson and Gaines, 1990). However, dispersal is selected for in some situations even when there is greater mortality of dispersing individuals, the optimal dispersal rate being inversely proportional to the mortality cost (Hamilton and May, 1977). It seems likely that a cost of dispersal would have changed the present results in that greater habitat fragmentation and extinction probability would have been required to select for dispersal, however, the general qualitative result would have remained unchanged.

Another assumption was that dispersal occurred spatially at random, i.e. there was no habitat selection. Morris (1991) showed that habitat selection was necessary to select for dispersal in a temporally constant environment, a similar result was obtained by McPeek and Holt (1992). These latter authors showed that a pure dispersal strategy with habitat selection is strongly selected in a temporally and spatially variable environment (McPeek and Holt, 1992). This is due to the strong advantage of those strategies able to track environmental changes. These results seem to be easily extended to the present study: habitat selection would have selected for dispersal in all situations. A similar result relative to the maintenance of dispersal polymorphism would also have been obtained by correctly tunig the parameters of habitat selection.

Other assumptions of the present model have been evaluated by simulations. The value of the mutation rate is important in that when it is decreased, unsurprisingly, the dynamics of the model are slower. This shows that the time scale of mutation has also a role in the dynamics of the model. Mutation is critical here as it allows the philopatric strategy to appear in other populations. For some organisms (e.g. insects), passive dispersal may play the same role than mutation in allowing the philopatric strategy to disperse in distant patches; such passive dispersal being much less frequent than active dispersal.

Most of the simulations were run assuming that only three dispersal strategies could exist. Assuming more strategies showed that in most cases an intermediate strategy was selected for. This suggests that for each combination of habitat fragmentation and temporal variation there exists an optimal dispersal rate. Assuming continuous variation in dispersal rate may not be a realistic for some organisms. Further informations on the range of possible dispersal rates seem necessary in order to use the present model for particular organisms.

The present model gives rise to predictions on the short-term evolution of dispersal. Selection for dispersal is expected to occur when the environment is fragmented or variable in time. To give testable predictions, the argument should be reversed: that is loss of dispersal abilities (for instance, brachyptery in insects) should be more frequent in spatially more continuous environments all other things being equal, or in temporally more stable environments all other things being equal.

**Acknowledgments**

I thank Derek Roff for useful comments on a previous version of the manuscript. Financial support for this study was provided by NERC grant GST/02/1197 as part of the NERC/SOAFED special topic on Large Scale Processes in Ecology and Hydrology.

*Rev. Ecol. Syst.* **14**, 159–187.

Figure 1. Evolution of the mean dispersal rate (denoted µ in the text) in a theoretical metapopulation (see the text for a detailed description of the model and the parameters), (a) the parameters are: $r_s = 1$, $r_k = -1$, $K_s = K_k = 1000$, $x_k = 1$, $x_s$ varies as indicated on the curves, (b) the parameters are: $x_s = x_k = 1$, $K_s = K_k = 1000$, $r_s = 1$, $r_k$ varies as indicated on the curves.

Figure 2. Mean dispersal rate in the metapopulation averaged along several time steps, (a) along the 2,500 time steps of the whole simulation, (b) along the last 500 time steps in order to remove possible transients effects.

Figure 3. Evolution of genotypic frequencies in the metapopulation with $x_s = 5$ and $p_e = 0.05$. Other parameters as in Fig. 1.

Figure 4. Same as in Fig. 3 except $p_e = 0.2$.

Figure 5. Same as in Fig. 3 except $p_e = 0.4$.

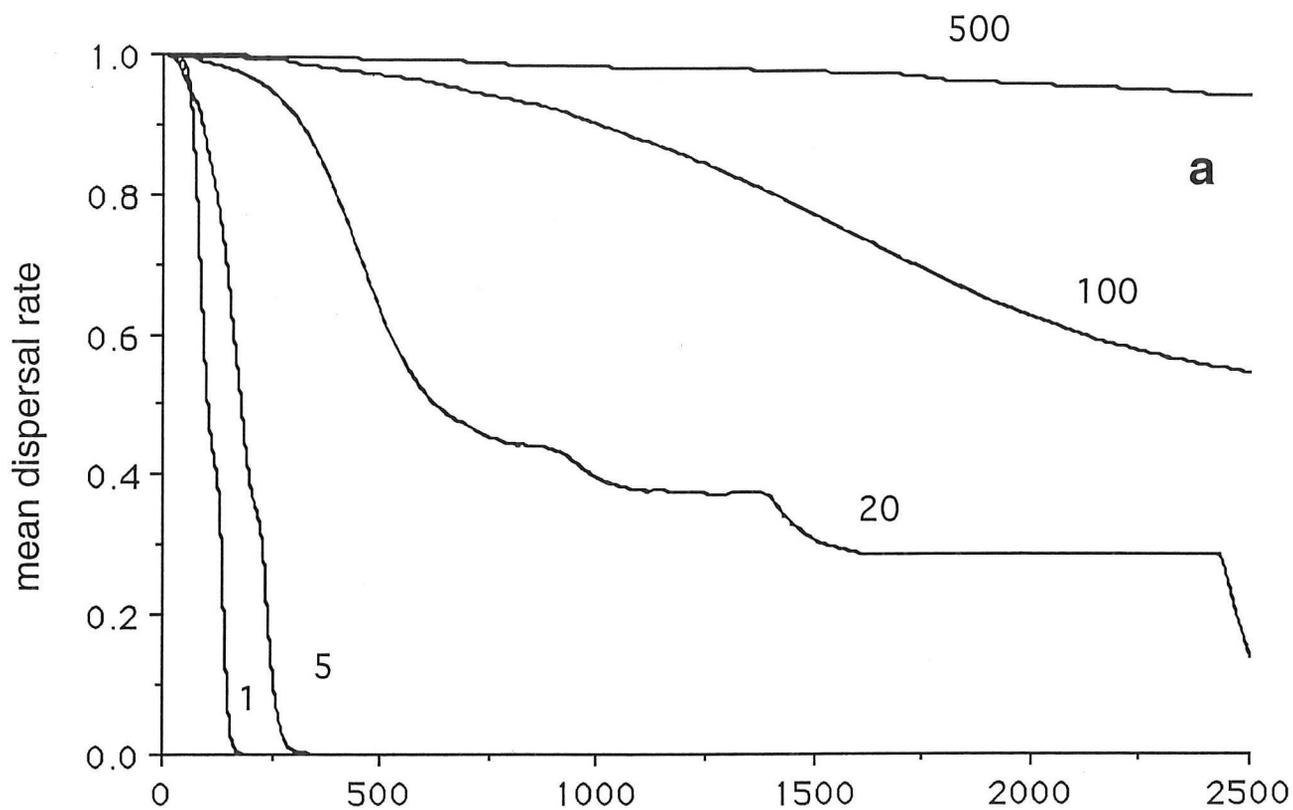

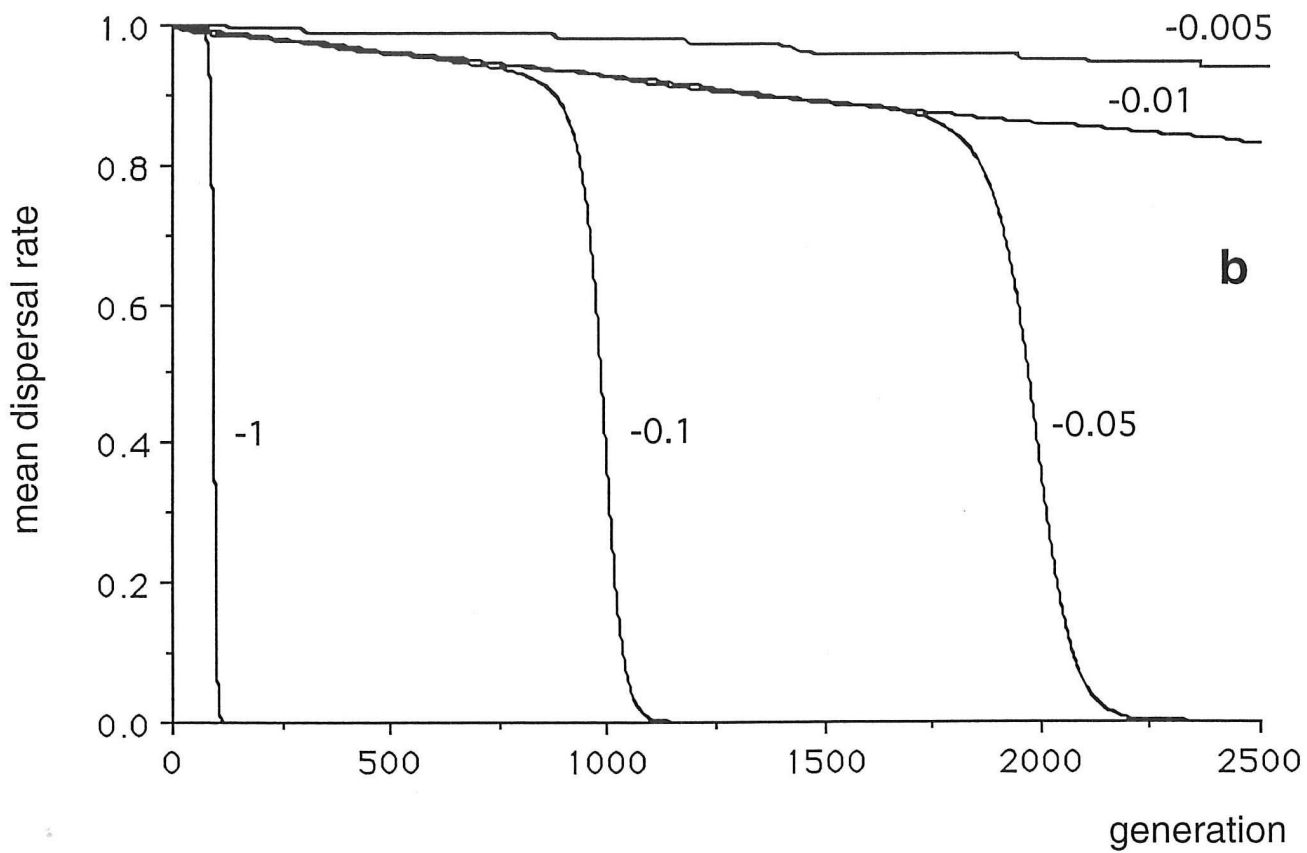

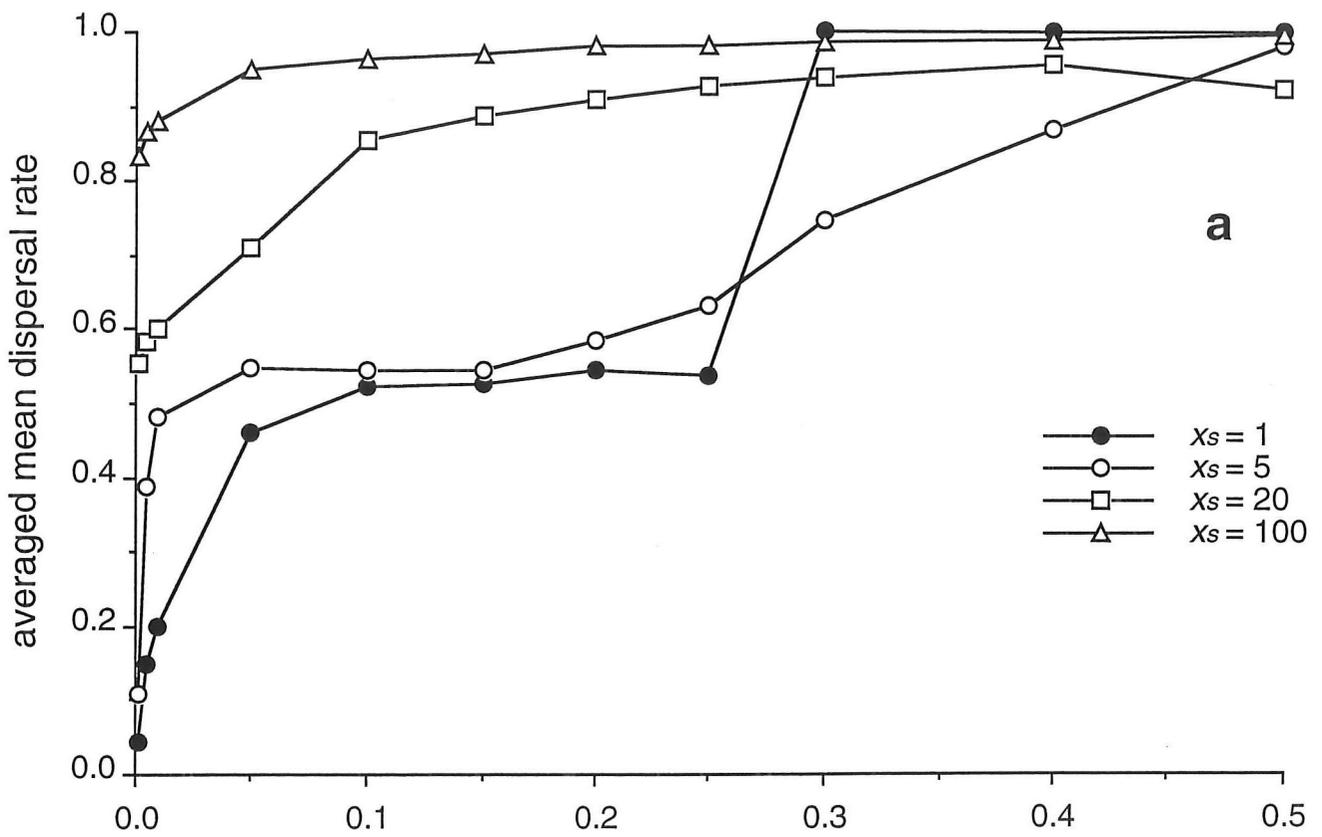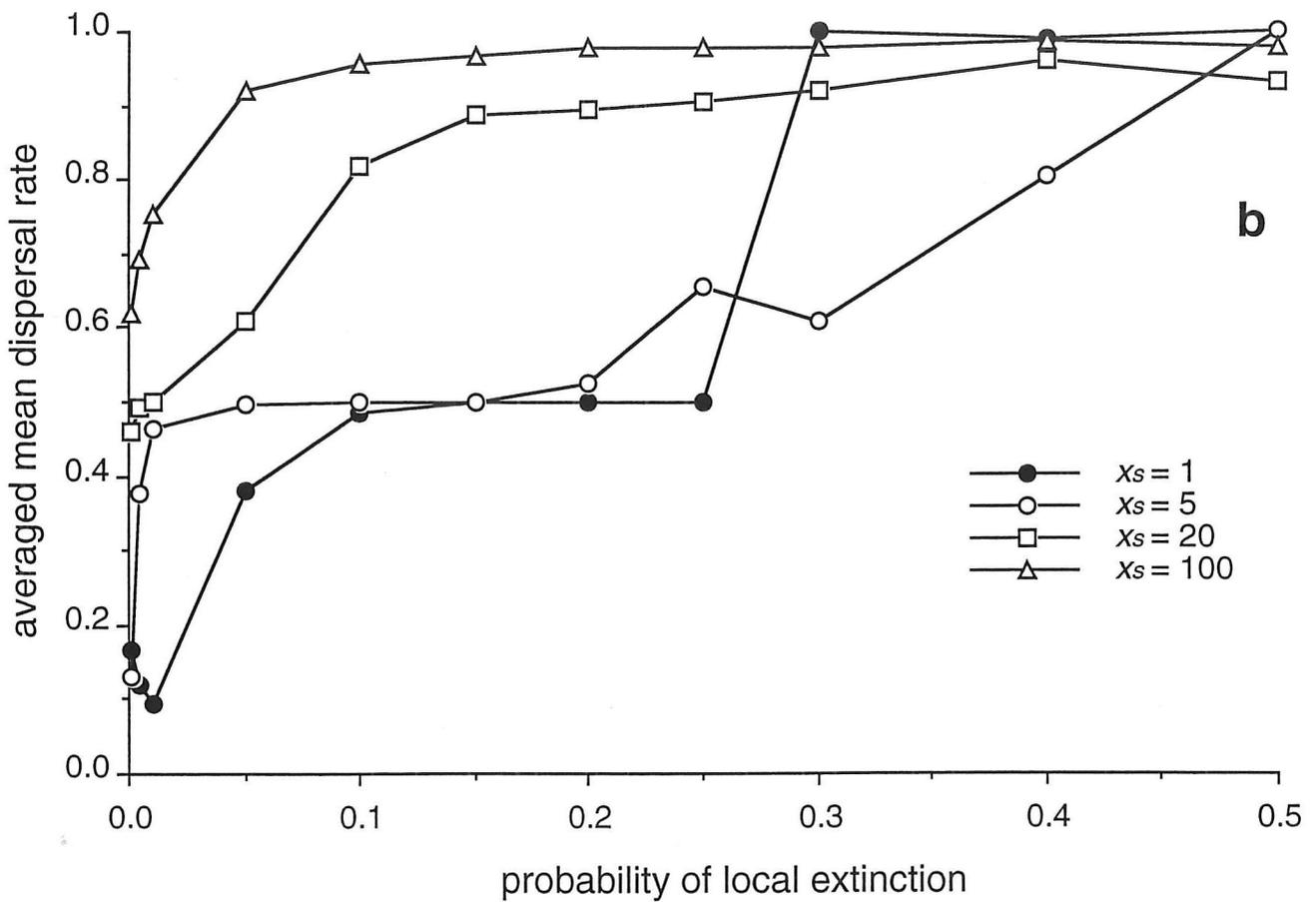

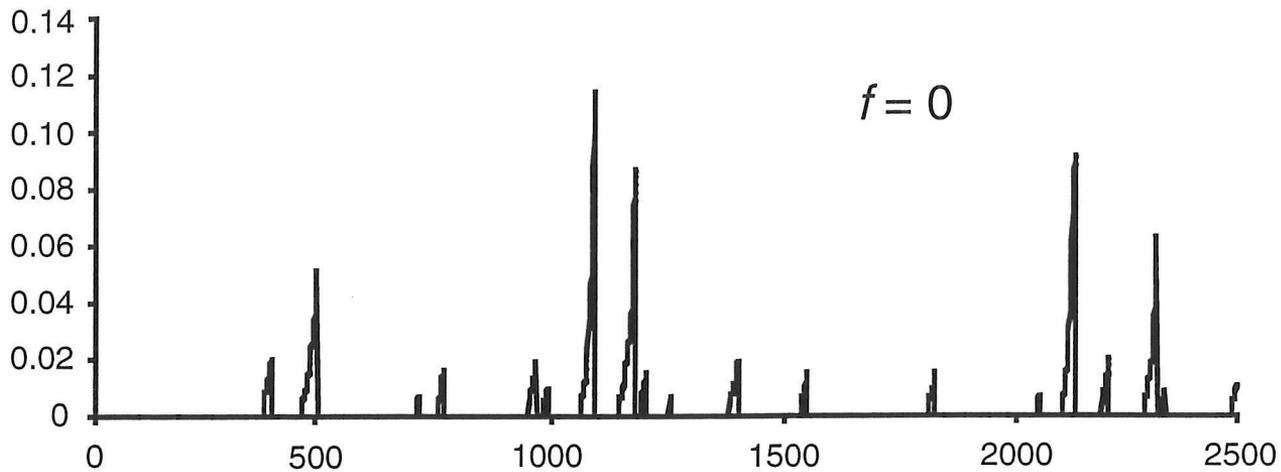
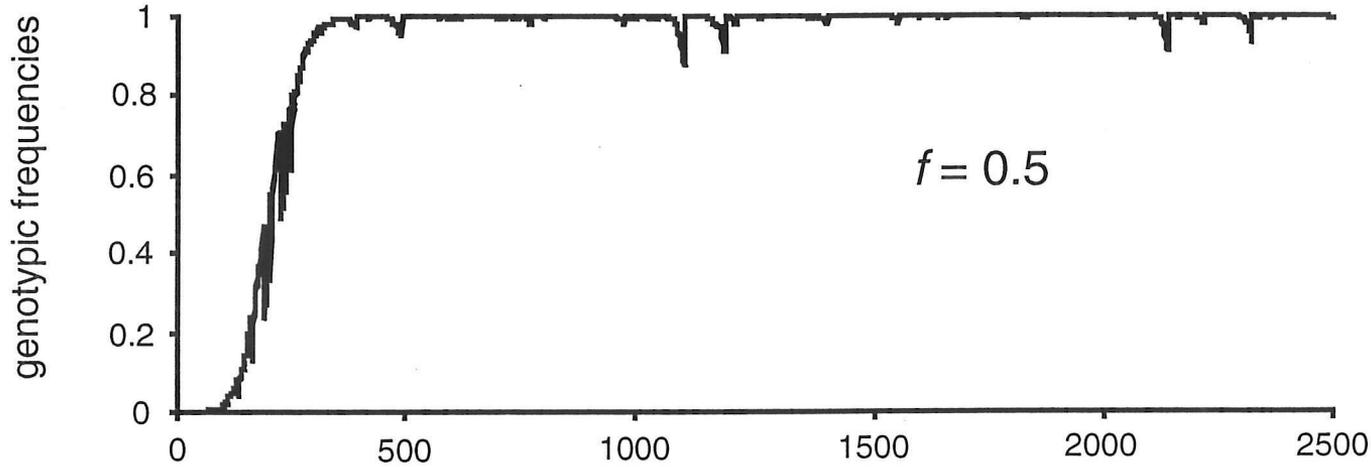
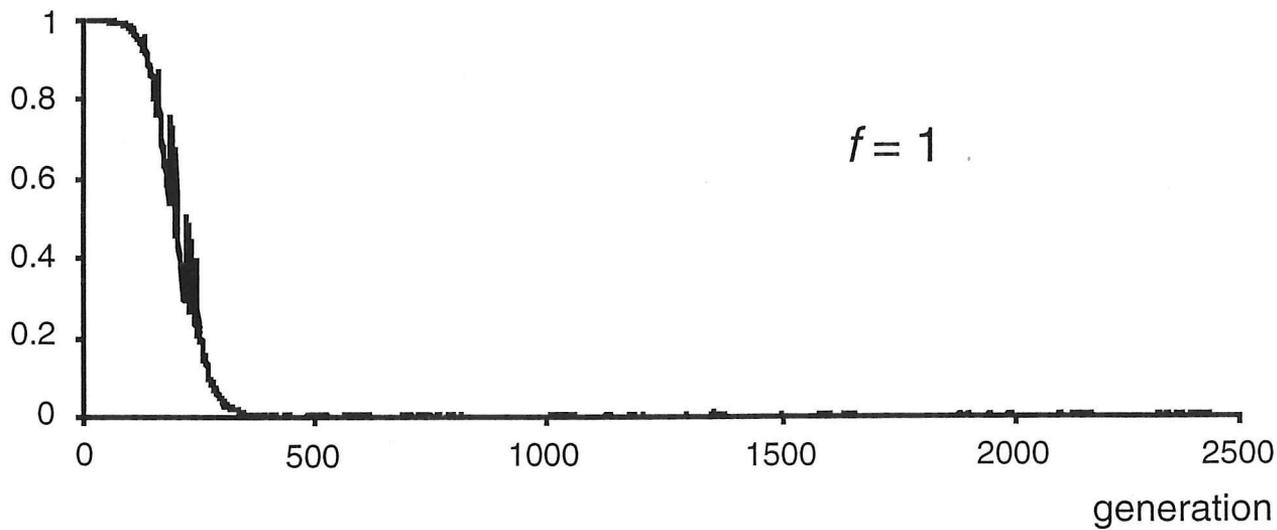

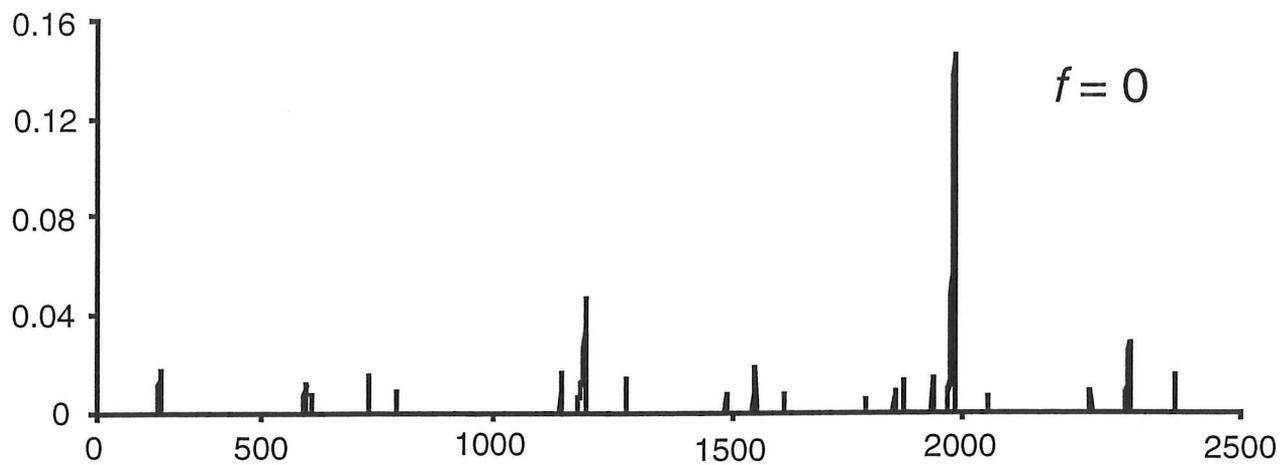
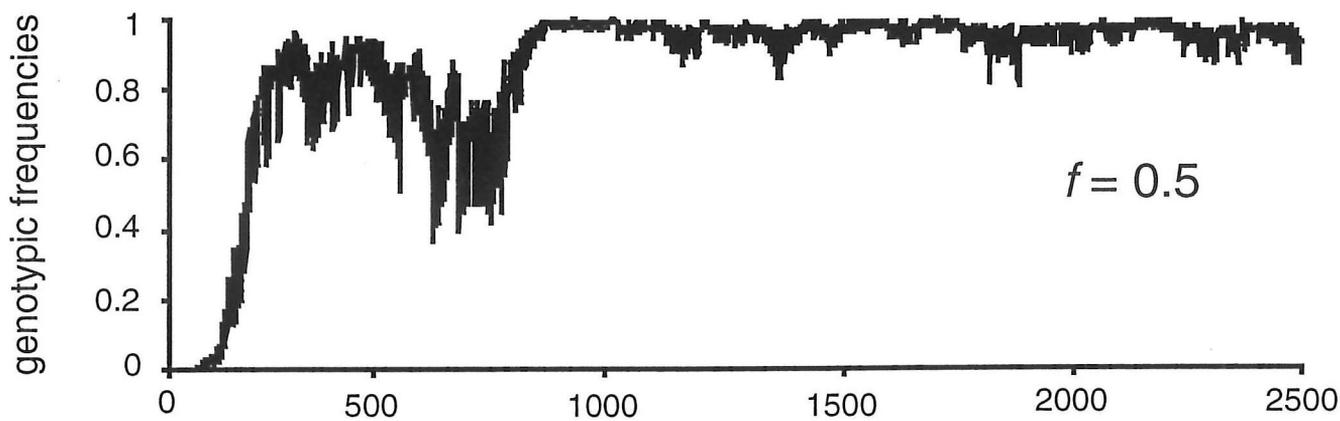
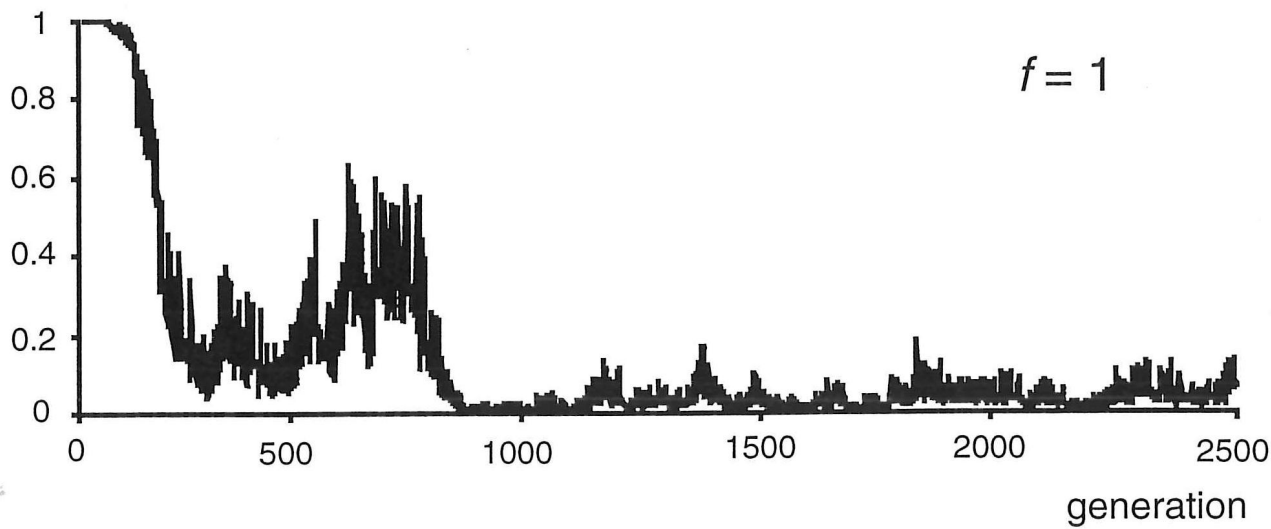

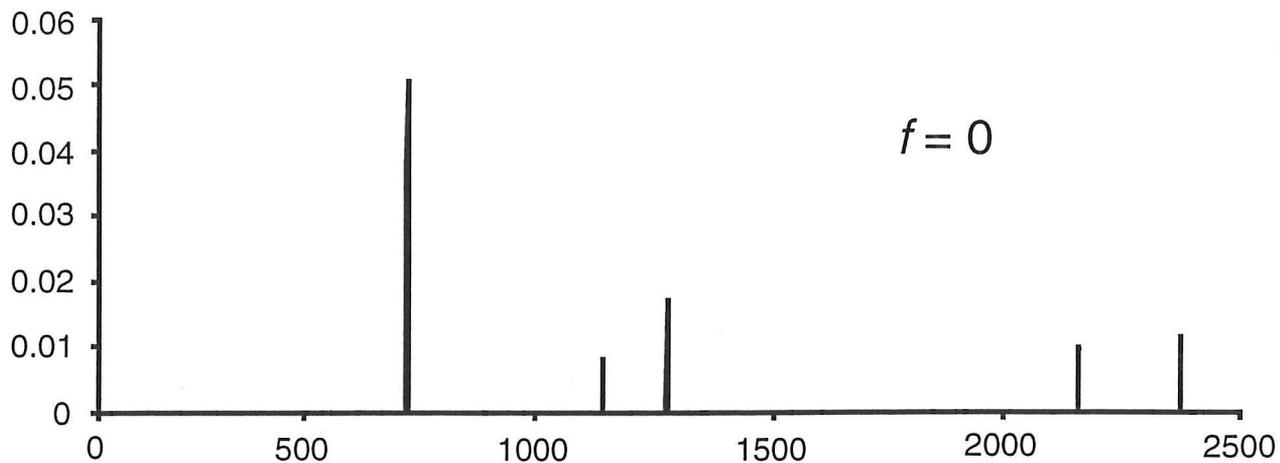

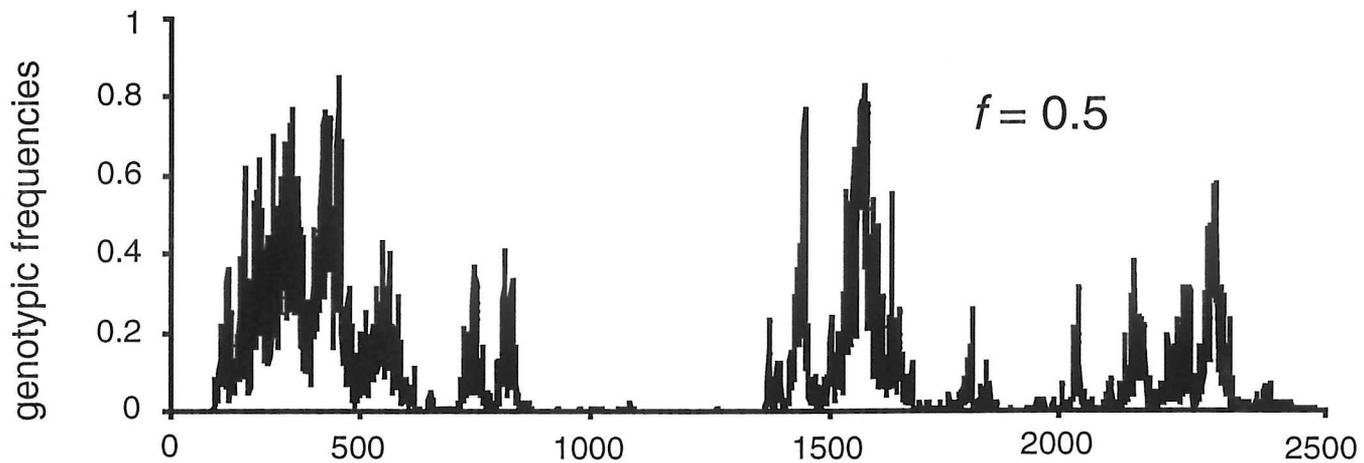

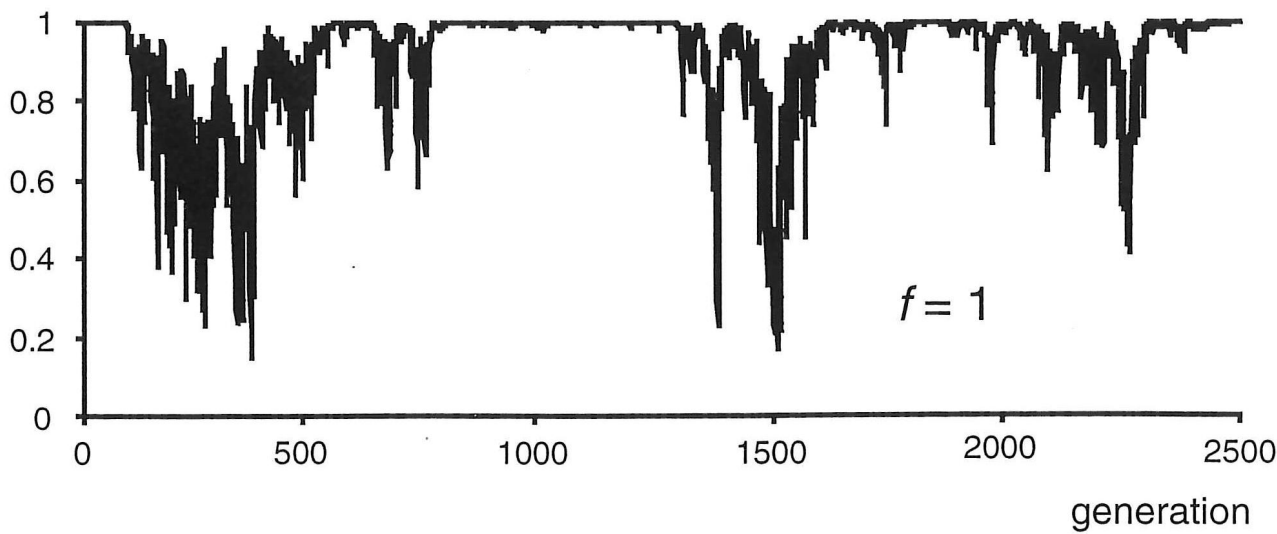